\documentclass[pra,aps]{revtex4}
\usepackage{graphicx}

\begin{document}

\title{Frustrated ferromagnetic spin-1/2 chain in the magnetic field}
\author{D.V.Dmitriev}
 \email{dmitriev@deom.chph.ras.ru}
\author{V.Ya.Krivnov}
\affiliation{Joint Institute of Chemical Physics of RAS, Kosygin
str.4, 119991, Moscow, Russia.}
\date{\today}

\begin{abstract}
We study the ground state properties of the Heisenberg spin-1/2 chain with
ferromagnetic nearest-neighbor and antiferromagnetic next-nearest-neighbor
interactions using two approximate methods. One of them is the Jordan-Wigner
mean-field theory and another approach is based on the transformation of
spin operators to bose-ones and on the variational treatment of bosonic
Hamiltonian. Both approaches give close results for the ground state energy
and the magnetization curve at $T=0$. It is proved that quantum fluctuations
change the classical critical exponents in the vicinity of the transition
point from the ferromagnetic to the singlet ground state. The magnetization
processes display a different behavior in the regions near and far from the
transition point. The relation of the obtained results to the experimental
magnetization curve in $Rb_{2}Cu_{2}Mo_{3}O_{12}$ is discussed.
\end{abstract}

\maketitle

\affiliation{Joint Institute of Chemical Physics of RAS, Kosygin
str.4, 119991, Moscow, Russia.}

\affiliation{Joint Institute of Chemical Physics of RAS, Kosygin
str.4, 119991, Moscow, Russia.}

\section{Introduction}

Lately, there has been considerable interest in low-dimensional quantum spin
systems that exhibit frustration. The spin-1/2 chain with
next-nearest-neighbor interactions (which is equivalent to a zig-zag ladder)
is a typical model with the frustration. The Hamiltonian of the model is
given by
\begin{equation}
H=\sum_{n=1}^{N}(J_{1}\mathbf{S}_{n}\cdot \mathbf{S}_{n+1}+J_{2}\mathbf{S}%
_{n}\cdot \mathbf{S}_{n+2})  \label{H}
\end{equation}
where $S$ is the spin-1/2 operator, $J_{1}$ and $J_{2}$ denote the nearest
(NN) and the next nearest neighbor (NNN) interactions.

This model with both NN and NNN antiferromagnetic interactions ($J_{1}$, $%
J_{2}>0$) is well studied \cite
{Bursill,Haldane,Majumdar,Okamoto,Tonegawa87,White}. There is a critical
value $J_{2c}=0.2411J_{1}$, which separates doubly degenerated dimer phase
(at $J_{2}>J_{2c}$) characterized by the excitation gap and the gapless
spin-fluid phase (at $J_{2}<J_{2c}$). Relatively less is known about the
model (\ref{H}) with the ferromagnetic NN and the antiferromagnetic NNN
interactions ($J_{1}<0,$ $J_{2}>0)$. Though the latter model has been a
subject of many studies \cite{Tonegawa89,Chubukov,Bursill,Cabra,KO} the
complete picture of the phases of this model remains unclear up to now. It
is known that the ground state is ferromagnetic for $J_{2}<\frac{\left|
J_{1}\right| }{4}$. At $J_{2}=\frac{\left| J_{1}\right| }{4}$ the
ferromagnetic state is degenerated with the singlet state. The wave function
of the singlet state at $J_{2}=\frac{\left| J_{1}\right| }{4}$ is known
exactly \cite{Hamada,DKO97}. For $J_{2}>\frac{\left| J_{1}\right| }{4}$ the
ground state is an incommensurate singlet. Controversial conclusions exist
about the presence of a gap at $J_{2}>\frac{\left| J_{1}\right| }{4}$. It
has been long believed that the model is gapless \cite{White,Allen} but the
one-loop renormalization group analysis indicates \cite{Nersesyan,Cabra}
that the gap is open due to a Lorentz symmetry breaking perturbation.
However, existence of the gap has not been verified numerically \cite{Cabra}%
. On the basis of a field theory consideration it has been proposed \cite
{Itoi} that a very tiny but finite gap exists which can be not observed
numerically.

Besides a general interest in frustrated quantum spin models there is an
additional motivation for the study of the model (\ref{H}) with $J_{1}<0$.
Really, the ferromagnetic NN interaction is expected to exist in compounds
containing $CuO$ chains with edge-sharing $CuO_{4}$ units. The $Cu-O-Cu$
angle in these compounds is close to $90^{0}$, and usual antiferromagnetic
NN exchange between $Cu$ ion spins is suppressed. This means that the sign
of $J_{1}$ can be negative, while the NNN exchange is antiferromagnetic.
Several compounds with edge-sharing chains are known \cite{Mizuno}, such as $%
Li_{2}CuO_{2},$ $La_{6}Ca_{8}Cu_{21}O_{41},$ $Ca_{2}Y_{2}Cu_{5}O_{10}$.
However, in these compounds the antiferromagnetic long range order appears
at low temperatures due to small interchain interactions. Recently, the
crystal $Rb_{2}Cu_{2}Mo_{3}O_{12}$ with edge-sharing chains has been
synthesized and studied experimentally \cite{Solodovnikov,Hase}. Remarkably,
no magnetic phase transition appears down to $2K$, which testifies that
interchain interaction is very small in this compound. Therefore, $%
Rb_{2}Cu_{2}Mo_{3}O_{12}$ can be considered as an ideal model compound with
the ferromagnetic (NN) interaction. According to \cite{Hase} it is described
by the Hamiltonian (\ref{H}) with $J_{1}\simeq -140K$ and $\frac{J_{2}}{%
\left| J_{1}\right| }\simeq 0.4$.

One of the interesting peculiarities of the model (\ref{H}) is the
dependence of the magnetization on the applied magnetic field at $T=0$. The
magnetization curve for $J_{1}<0$ is quite different from that for the case $%
J_{1}>0$. It is characterized by a rapid increase (or even discontinuity) in
the magnetization, if the external field exceeds a critical value. It is
expected that the magnetization displays a true jump (the metamagnetic
transition) when the NNN interaction $J_{2}$ is slightly larger than $\frac{%
\left| J_{1}\right| }{4}$ \cite{Aligia}. This conclusion has been made on
the basis of the exact diagonalization calculations for finite chains.
However, the extrapolation of these results to the thermodynamic limit is
rather difficult due to strong non-monotonic finite size effects.

In this paper the ground state energy and the magnetization processes in the
model (\ref{H}) with $J_{1}<0$ are studied using two variational approaches.
One of these approaches is based on the Jordan-Wigner transformation of
spin-1/2 operators to the Fermi ones with the subsequent mean-field
treatment of the Fermi-Hamiltonian. Another variational approach is applied
to the Bose-Hamiltonian arising from a special transformation of spin
operators to Bose ones.

The paper is organized as follows. In Section II the details of the
Jordan-Wigner mean-field approximation are presented. The zero-temperature
magnetization process is studied and the form of the magnetization curve in
different region of the parameter $\frac{J_{2}}{\left| J_{1}\right| }$ is
found. For the case $h=0$ we have focused on the behavior of the model in
the vicinity of the transition point from the ferromagnetic to the singlet
ground state. The critical exponents characterizing this behavior are
determined. In Section III we describe the variational method for the
treatment of the considered Hamiltonian rewritten in bosonic form. Scaling
estimates of the critical exponent near the transition point are presented
in Section IV. In Section V we summarize our results.

\section{Jordan-Wigner mean-field approach}

It is convenient to represent the Hamiltonian of the spin-1/2 chain with the
ferromagnetic NN and the antiferromagnetic NNN interactions in the form:
\begin{equation}
H=-\sum_{n=1}^{N}(\mathbf{S}_{n}\cdot \mathbf{S}_{n+1}-\frac{1}{4})+\alpha
\sum_{n=1}^{N}(\mathbf{S}_{n}\cdot \mathbf{S}_{n+2}-\frac{1}{4}%
)-h\sum_{n=1}^{N}S_{n}^{y}  \label{H1}
\end{equation}
where $\alpha =\frac{J_{2}}{\left| J_{1}\right| }$ and $h=\frac{g\mu _{B}B}{%
\left| J_{1}\right| }$ is the effective dimensionless magnetic field. The
constant shifts in Eq.(\ref{H1}) secures the energy of the fully polarized
state to be zero.

Let us start from the classical picture of the ground state of Eq.(\ref{H1}%
). In the classical approximation the spins are vectors which form the
spiral structure in the $XZ$ plane with a pitch angle $\varphi $ between
neighboring spins and all spin vectors are inclined towards the $Y$ axis by
an angle $\theta $. The classical energy per site
\begin{equation}
\epsilon (\varphi ,\theta )=\frac{1}{4}\left[ 1-\cos \varphi -\alpha (1-\cos
(2\varphi )\right] \cos ^{2}\theta -\frac{h}{2}\sin \theta
\end{equation}
is minimized by the angles
\begin{equation}
\varphi _{cl}=\cos ^{-1}\frac{1}{4\alpha },\text{ \ }\theta _{cl}=\sin ^{-1}%
\frac{\alpha h}{2\gamma ^{2}}
\end{equation}
where $\gamma =\alpha -\frac{1}{4}$.

The classical ground state energy is
\begin{equation}
\frac{E_{cl}}{N}=-\frac{1}{2\alpha }\gamma ^{2}-\frac{\alpha h^{2}}{8\gamma
^{2}}
\end{equation}

Following this picture we transform local axes on $n$-th site by a rotation
about the $Y$ axis by $\varphi n$ and then by a rotation about the $X$ axis
by $\theta $. The transformation to new spin-$\frac{1}{2}$ operators $%
\mathbf{\eta }_{n}$ has a form
\begin{equation}
\mathbf{S}_{n}=R_{y}(\varphi n)R_{x}(\theta )\mathbf{\eta }_{n}
\label{rotate}
\end{equation}
where $R_{y}(\varphi n)$ and $R_{x}(\theta )$ are the operators of the
corresponding rotations.

Substituting (\ref{rotate}) into (\ref{H1}) we obtain the transformed
Hamiltonian in terms of the $\mathbf{\eta }$ operators
\begin{eqnarray}
H &=&H_{1}+H_{2}+H_{3}  \label{Hrot} \\
H_{1} &=&N\epsilon (\varphi ,\theta )+\sum_{n=1}^{N}\left[ J_{1x}\eta
_{n}^{x}\eta _{n+1}^{x}+J_{1y}\eta _{n}^{y}\eta _{n+1}^{y}+J_{1z}(\eta
_{n}^{z}\eta _{n+1}^{z}-\frac{1}{4})\right] +  \nonumber \\
&&+\sum_{n=1}^{N}\left[ J_{2x}\eta _{n}^{x}\eta _{n+2}^{x}+J_{2y}\eta
_{n}^{y}\eta _{n+2}^{y}+J_{2z}(\eta _{n}^{z}\eta _{n+2}^{z}-\frac{1}{4})%
\right] -h\sin \theta \sum_{n=1}^{N}(\eta _{n}^{z}-\frac{1}{2})  \nonumber \\
H_{2} &=&\sin \theta \sum_{n=1}^{N}[\sin \varphi (\eta _{n}^{x}\eta
_{n+1}^{y}-\eta _{n}^{y}\eta _{n+1}^{x})-\alpha \sin 2\varphi (\eta
_{n}^{x}\eta _{n+2}^{y}-\eta _{n}^{y}\eta _{n+2}^{x})]  \nonumber \\
H_{3} &=&-\sum_{n=1}^{N}[\frac{\sin 2\theta }{2}(1-\cos \varphi )(\eta
_{n}^{z}\eta _{n+1}^{y}+\eta _{n}^{y}\eta _{n+1}^{z})-\alpha \frac{\sin
2\theta }{2}(1-\cos 2\varphi )(\eta _{n}^{z}\eta _{n+2}^{y}+\eta
_{n}^{y}\eta _{n+2}^{z})  \nonumber \\
&&+\sin \varphi \cos \theta (\eta _{n}^{x}\eta _{n+1}^{z}-\eta _{n}^{z}\eta
_{n+1}^{x})-\alpha \sin 2\varphi \cos \theta (\eta _{n}^{x}\eta
_{n+2}^{z}-\eta _{n}^{z}\eta _{n+2}^{x})]-h\cos \theta \sum_{n=1}^{N}\eta
_{n}^{y}  \nonumber
\end{eqnarray}
where
\begin{eqnarray}
J_{1x} &=&-\cos \varphi ,\qquad J_{1y}=-\cos \varphi \sin ^{2}\theta -\cos
^{2}\theta ,\qquad J_{1z}=-\cos \varphi \cos ^{2}\theta -\sin ^{2}\theta
\nonumber \\
J_{2x} &=&\alpha \cos 2\varphi ,\qquad J_{2y}=\alpha (\cos 2\varphi \sin
^{2}\theta +\cos ^{2}\theta ),\qquad J_{2z}=\alpha (\cos \varphi \cos
^{2}\theta +\sin ^{2}\theta )  \label{Jxyz}
\end{eqnarray}

The Hamiltonian (\ref{Hrot}) for $h=0$ ($\theta =0$) has been considered
before in \cite{KO,Bursill}. In Ref.\cite{KO} the ground state in the
vicinity of the point $\alpha =\frac{1}{4}$ has been studied on the basis of
the spin-wave theory taking $\varphi $ by its classical value. It was shown
that the transition at $\alpha =\frac{1}{4}$ from the ferromagnetic state to
the singlet one is of the second order and the ground state energy is $%
-4\gamma ^{2}$ for $0<\gamma \ll 1$. In \cite{Bursill} the dependence of the
ground state energy and the pitch angle $\varphi $ on $\alpha $ has been
found using the coupled cluster method. In particular, the ground state
energy is proportional to $\gamma ^{2}$ for $0<\gamma \ll 1$ too. However,
both cited approaches are not variational ones. Besides, the magnetization
curve has not been studied.

Our primary interest is how quantum effects alter the classical ground state
structure. In this section we use the Jordan-Wigner transformation which
converts the spin Hamiltonian (\ref{Hrot}) into a model of interacting
spinless fermions
\begin{equation}
\eta _{n}^{+}=c_{n}\exp (i\pi \sum_{j=1}^{n-1}c_{j}^{+}c_{j}),\text{ \ \ \ \
}\eta _{n}^{z}=\frac{1}{2}-c_{n}^{+}c_{n}\text{\ \ \ }  \label{JW}
\end{equation}

The Hamiltonians $H_{1}$ and $H_{2}$ in Eq.(\ref{Hrot}) are transformed to
the Fermi-Hamiltonians having forms:
\begin{eqnarray}
H_{1f} &=&N\epsilon (\varphi ,\theta )+(h\sin \theta
-J_{1z}-J_{2z})\sum_{n=1}^{N}c_{n}^{+}c_{n}+J_{1z}%
\sum_{n=1}^{N}c_{n}^{+}c_{n}c_{n+1}^{+}c_{n+1}+J_{2z}%
\sum_{n=1}^{N}c_{n}^{+}c_{n}c_{n+2}^{+}c_{n+2}  \nonumber \\
&&+\frac{1}{4}(J_{1x}+J_{1y})%
\sum_{n=1}^{N}(c_{n}^{+}c_{n+1}+c_{n+1}^{+}c_{n})+\frac{1}{4}%
(J_{2x}+J_{2y})%
\sum_{n=1}^{N}(c_{n}^{+}c_{n+2}+c_{n+2}^{+}c_{n})(1-2c_{n+1}^{+}c_{n+1})
\nonumber \\
&&+\frac{1}{4}(J_{1x}-J_{1y})%
\sum_{n=1}^{N}(c_{n}^{+}c_{n+1}^{+}+c_{n+1}c_{n})+\frac{1}{4}%
(J_{2x}-J_{2y})%
\sum_{n=1}^{N}(c_{n}^{+}c_{n+2}^{+}+c_{n+2}c_{n})(1-2c_{n+1}^{+}c_{n+1})
\label{Hf} \\
H_{2f} &=&\frac{i}{2}\sin \theta \sum_{n=1}^{N}[\sin \varphi
(c_{n}^{+}c_{n+1}-c_{n+1}^{+}c_{n})-\alpha \sin 2\varphi
(c_{n}^{+}c_{n+2}-c_{n+2}^{+}c_{n})(1-2c_{n+1}^{+}c_{n+1})]  \nonumber
\end{eqnarray}

We do not present here the Hamiltonian $H_{3f}$ because it has very
complicated form containing non-local interaction and, as it will be shown
below, does not contribute to the energy in the mean-field approach.

The next step of the approach is to treat the Hamiltonian $%
H_{f}=H_{1f}+H_{2f}+H_{3f}$ in the mean-field approximation (MFA). We use
the approximation scheme proposed in \cite{DK}. The ground state wave
function in this approximation has BCS-like form
\[
\left| \Psi \right\rangle =\prod_{k>0}(\cos \phi _{k}+\sin \phi
_{k}c_{k}^{+}c_{-k}^{+})\left| 0\right\rangle
\]
where $c_{k}^{+}$ is Fourier transform of the Fermi-operators $c_{n}^{+}$.

The expectation value of $H_{2f}$ over the function $\Psi $ is $\left\langle
H_{2f}\right\rangle =0$ due to the fact that $\sum_{k}\sin k\left\langle
c_{k}^{+}c_{k}\right\rangle =0$\ and $\sum_{k}\cos k\left\langle
c_{k}^{+}c_{-k}^{+}\right\rangle =0$. The Hamiltonian $H_{3f}$ includes odd
number of the Fermi-operators and, therefore, it is zero in the MFA $%
\left\langle H_{3f}\right\rangle =0$ as well. Thus, both Hamiltonians $%
H_{2f} $ and $H_{3f}$ do not contribute to the energy in this approximation.

In the MFA four-fermions terms in Eqs.(\ref{Hf}) are decoupled in all
possible ways. After Fourier transformation the mean-field Hamiltonian takes
a form
\begin{equation}
H_{\mathrm{MFA}}=NC+\sum_{k>0}a(k)(c_{k}^{+}c_{k}+c_{-k}^{+}c_{-k})+%
\sum_{k>0}b(k)(c_{k}^{+}c_{-k}^{+}+c_{-k}c_{k})  \label{Hmfa}
\end{equation}
where
\begin{eqnarray}
C &=&\epsilon (\varphi ,\theta
)-(J_{1z}+J_{2z})n^{2}+J_{1z}(t_{1}^{2}-g_{1}^{2})+J_{2z}(t_{2}^{2}-g_{2}^{2})+
\label{C} \\
&&(J_{2x}+J_{2y})(nt_{2}-t_{1}^{2}-g_{1}^{2})+(J_{2x}-J_{2y})(ng_{2}-2t_{1}g_{1})
\nonumber
\end{eqnarray}
\begin{eqnarray}
a(k) &=&u+\nu _{1}\cos k+\nu _{2}\cos 2k  \label{ab} \\
b(k) &=&w_{1}\sin k+w_{2}\sin 2k  \nonumber
\end{eqnarray}
\begin{eqnarray}
u &=&-(J_{1z}+J_{2z})(1-2n)-(J_{2x}+J_{2y})t_{2}-(J_{2x}-J_{2y})g_{2}+h\sin
\theta  \label{uvw} \\
\nu _{1} &=&\frac{1}{2}%
(J_{1x}+J_{1y})-2J_{1z}t_{1}+2(J_{2x}+J_{2y})t_{1}+2(J_{2x}-J_{2y})g_{1}
\nonumber \\
\nu _{2} &=&(J_{2x}+J_{2y})(\frac{1}{2}-n)-2J_{2z}t_{2}  \nonumber \\
w_{1} &=&\frac{1}{2}%
(J_{1x}-J_{1y})+2(J_{1z}+J_{2x}+J_{2y})g_{1}+2(J_{2x}-J_{2y})t_{1}  \nonumber
\\
w_{2} &=&(J_{2x}-J_{2y})(\frac{1}{2}-n)+2J_{2z}g_{2}  \nonumber
\end{eqnarray}

The ground state energy, the one-particle excitation spectrum $\varepsilon
(k)$\ and the magnetization $M=\langle S_{n}^{y}\rangle $\ are:
\begin{eqnarray}
\frac{E}{N} &=&\epsilon (\varphi ,\theta )+F(\varphi ,\theta )  \nonumber \\
\varepsilon (k) &=&\sqrt{a^{2}(k)+b^{2}(k)}  \nonumber \\
M &=&\left( \frac{1}{2}-n\right) \sin \theta
\end{eqnarray}
where
\begin{eqnarray}
F(\varphi ,\theta )
&=&(J_{1z}+J_{2z})n(1-n)+J_{1z}(g_{1}^{2}-t_{1}^{2})+J_{2z}(g_{2}^{2}+t_{2}^{2}-2t_{2})+%
\frac{1}{2}(J_{1x}+J_{1y})t_{1}  \nonumber \\
&&+\frac{1}{2}%
(J_{1x}-J_{1y})g_{1}+(J_{2x}+J_{2y})(nt_{2}-t_{1}^{2}-g_{1}^{2})+(J_{2x}-J_{2y})(ng_{2}-2t_{1}g_{1})
\end{eqnarray}

Quantities $n,t_{1},t_{2},g_{1},g_{2}$ are the ground state expectation
values, which are determined by the self-consistency equations:
\begin{eqnarray}
n &=&\langle c_{n}^{+}c_{n}\rangle =\int\limits_{0}^{\pi }\frac{dk}{2\pi }%
\left( 1-\frac{a(k)}{\varepsilon (k)}\right)  \nonumber \\
t_{1} &=&\langle c_{n}^{+}c_{n+1}\rangle =-\int\limits_{0}^{\pi }\frac{dk}{%
2\pi }\frac{a(k)\cos k}{\varepsilon (k)}  \nonumber \\
t_{2} &=&\langle c_{n}^{+}c_{n+2}\rangle =-\int\limits_{0}^{\pi }\frac{dk}{%
2\pi }\frac{a(k)\cos 2k}{\varepsilon (k)}  \nonumber \\
g_{1} &=&\langle c_{n}^{+}c_{n+1}^{+}\rangle =-\int\limits_{0}^{\pi }\frac{dk%
}{2\pi }\frac{b(k)\sin k}{\varepsilon (k)}  \nonumber \\
g_{2} &=&\langle c_{n}^{+}c_{n+2}^{+}\rangle =-\int\limits_{0}^{\pi }\frac{dk%
}{2\pi }\frac{b(k)\sin 2k}{\varepsilon (k)}  \label{ntq}
\end{eqnarray}

The solution of the self-consistency equations gives the minimum of the
ground state energy in a class of a `one-particle' wave functions at given
angles $\varphi $ and $\theta $. We treat the angles $\varphi $ and $\theta $
as the variational parameters (not equal to their classical values). Thus,
one should minimize the energy with respect to the angles $\varphi $ and $%
\theta $, solving the self-consistency equations for each value of $\varphi $
and $\theta $. This means that the proposed procedure remains variational
one.

To study the effect of the dimerization we added the staggered terms to the
NN expectation values:
\begin{eqnarray*}
\langle c_{n}^{+}c_{n+1}\rangle &=&t_{1}+t_{1}^{\prime }(-1)^{n} \\
\langle c_{n}^{+}c_{n+1}^{+}\rangle &=&g_{1}+g_{1}^{\prime }(-1)^{n}
\end{eqnarray*}
which leads to more complicated form of the Hamiltonian $H_{\mathrm{MFA}}$
and two additional self-consistency equations for $t_{1}^{\prime }$ and $%
g_{1}^{\prime }$. But the solution of the self-consistency equations with
the minimization of the energy over $\varphi $ and $\theta $ gives $%
t_{1}^{\prime }=g_{1}^{\prime }=0$. This means that the MFA does not
indicate the dimerization in the system.

\subsection{Ground state energy near the transition point $\protect\alpha %
=\frac 14$}

At first we consider the Hamiltonian (\ref{Hmfa}) at $h=0$ when $\theta =0$
and we are interested mainly in the behavior of the model in the vicinity of
the point $\alpha =\frac 14$, where the transition from the ferromagnetic to
the singlet ground state occurs. At $h=0$ the ground state energy per site
is:
\begin{equation}
\frac{E}{N}=\epsilon (\varphi ,0)+F(\varphi ,0)  \label{EepsF}
\end{equation}
where $F(\varphi ,0)$ is a quantum correction to the classical part of the
energy.

The analysis of the solution of Eqs.(\ref{ntq}) shows that in the case $%
0<\gamma \ll 1$ and $\varphi \ll 1$ the functions $\epsilon (\varphi ,0)$
and $F(\varphi ,0)$ have forms:
\begin{eqnarray}
\epsilon (\varphi ,0) &=&-\frac{\gamma \varphi ^{2}}{2}+\frac{\varphi ^{4}}{%
32}  \nonumber \\
F(\varphi ,0) &=&-\frac{\varphi ^{4}}{32}+A\varphi ^{\frac{24}{5}}
\label{Fexpan}
\end{eqnarray}

The coefficient $A\approx 0.0195$ is determined by the numerical solution of
Eqs.(\ref{ntq}).

Substituting (\ref{Fexpan}) into (\ref{EepsF}) and minimizing $E$ with
respect to $\varphi $ we obtain the leading terms in $\gamma $ for the angle
$\varphi (\gamma )$ and the energy $E(\gamma )$:
\begin{eqnarray}
\varphi &=&2.331\gamma ^{\frac{5}{14}}  \nonumber \\
\frac{E}{N} &=&-1.585\gamma ^{\frac{12}{7}}  \label{Egamma}
\end{eqnarray}

\begin{figure}[tbp]
\includegraphics{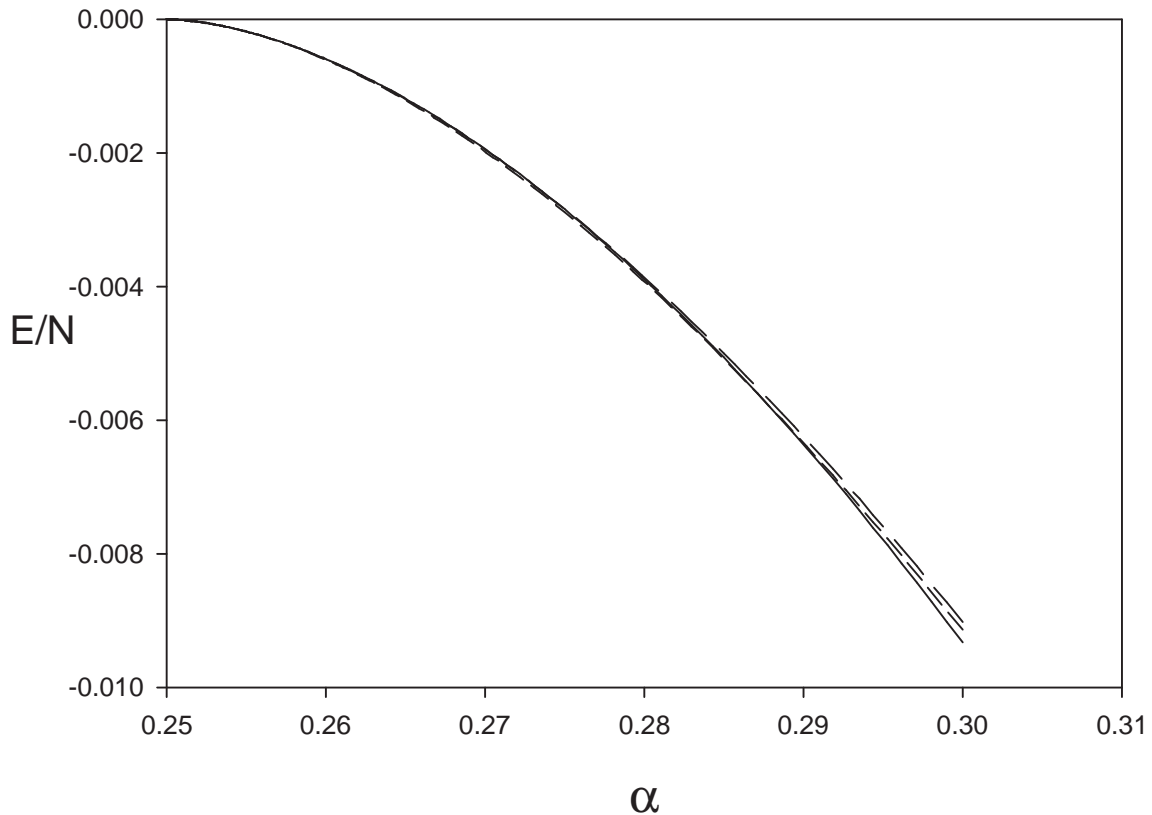}
\caption{The dependence of the ground state energy of the model (\ref{H1})
on $\protect\alpha $ in the MFA (short-dashed line) and the boson
variational approach (long-dashed line). Solid line corresponds to the
energy given by Eq.(\ref{Egamma}).}
\label{fig_1}
\end{figure}

The numerical solution of self-consistency equations (\ref{ntq}) confirms
this dependence $E(\gamma )$ at $0<\gamma \ll 1$ (see Fig.\ref{fig_1}).

Since the MFA is the variational approach, the found critical exponent for
the ground state energy $\beta =\frac{12}{7}$\ gives upper bound for the
exact critical exponent $\beta \leq \frac{12}{7}$\ and certainly it is less
than $\beta =2$\ obtained in classical approximation. We note that the use
of more elaborate methods \cite{KO,Bursill} does not change the exponent $%
\beta =2$\ and change the numerical prefactor $a$\ in the ground state
energy $E=-a\gamma ^{2}$\ only. This means that at present the MFA gives the
best estimate of the ground state energy in the region $0<\gamma \ll 1$.

In Ref.\cite{Sun} the model (\ref{H1}) (at $h=0$) has been studied using the
Jordan-Wigner transformation and the mean-field theory. However, the pair
correlations of type $\langle c_{n}^{+}c_{m}^{+}\rangle $ have been
neglected. It was conjectured in \cite{Sun} that the ground state at $\frac{1%
}{4}<\alpha <0.25854$ is not singlet but has a nonzero total spin. We have
found no evidence of this fact and the numerical diagonalization of finite
chains does not confirm this prediction as well.

As was written above, in our mean-field treatment the pitch angle $\varphi $
is considered as the variational parameter. It should be noted that in the
MFA a helical (spiral) long range order (LRO) exists as well as in the
classical approximation. The helical LRO is characterized by the angle $%
\varphi $, though its value differs from the classical one. Of course, this
LRO is destroyed by those quantum fluctuations\ which are ignored in the
MFA. At the same time the obtained helical LRO is regarded to an
incommensurate behavior of the correlation function and $\varphi $ can be
identified with the momentum $q_{\max }$ at which the static structure
factor takes a maximal value.

\subsection{Magnetization curve}

In this section we consider the magnetization processes in the model (\ref
{H1}). The behavior of the magnetization in the region of the magnetic field
close to saturation is of a particular interest. We note that the
determination of the saturation field $h_{s}(\alpha )$ at which the
transition to saturation takes place is generally not a simple problem. In
the model (\ref{H}) with both antiferromagnetic NN and NNN interactions this
field is equal to the energy of an one-magnon state on the ferromagnetic
background (all spins up), i.e. $h_{s}=-E_{1}$. It is not the case for the
model (\ref{H1}). The specific property of this model is the presence of
bound magnon states (complexes of two or more flipped spins) arising as a
result of magnon-magnon attraction. When the field $h$\ is reduced below the
saturated value the number of spins flipping simultaneously is more than
one. Then the saturated field is determined by the condition
\begin{equation}
h_{s}=\max \left\{ \frac{\left| E_{n}\right| }{n}\right\}  \label{hc}
\end{equation}
where $E_{n}$ is the energy of $n$-magnon state.

There are two possible scenarios of the behavior of the magnetization curve
close to saturation. In the first case Eq.(\ref{hc}) is satisfied at $%
n^{\ast }\sim o(N)$ and the magnetization is a continuous function at $%
h\rightarrow h_{s}$. In the second case $n^{\ast }\sim N$ and the
magnetization jumps at $h=h_{s}$ from $M^{\ast }=(\frac{1}{2}-\frac{n^{\ast }%
}{N})$ to $M=\frac{1}{2}$, i.e. the metamagnetic transition occurs. It means
that the ground state energy as a function of magnetization $e(M)$ has a
negative curvature and $M^{\ast }$ is determined by the Maxwell
construction. We note that the numerical estimate of $h_{s}$ and $n^{\ast }$
especially at $0<\gamma \ll 1$ is very difficult because the number of spins
flipping simultaneously at $M\rightarrow \frac{1}{2}$ increases when $\alpha
\rightarrow \frac{1}{4}$ \cite{Cabra,Aligia}.

\begin{figure}[tbp]
\includegraphics{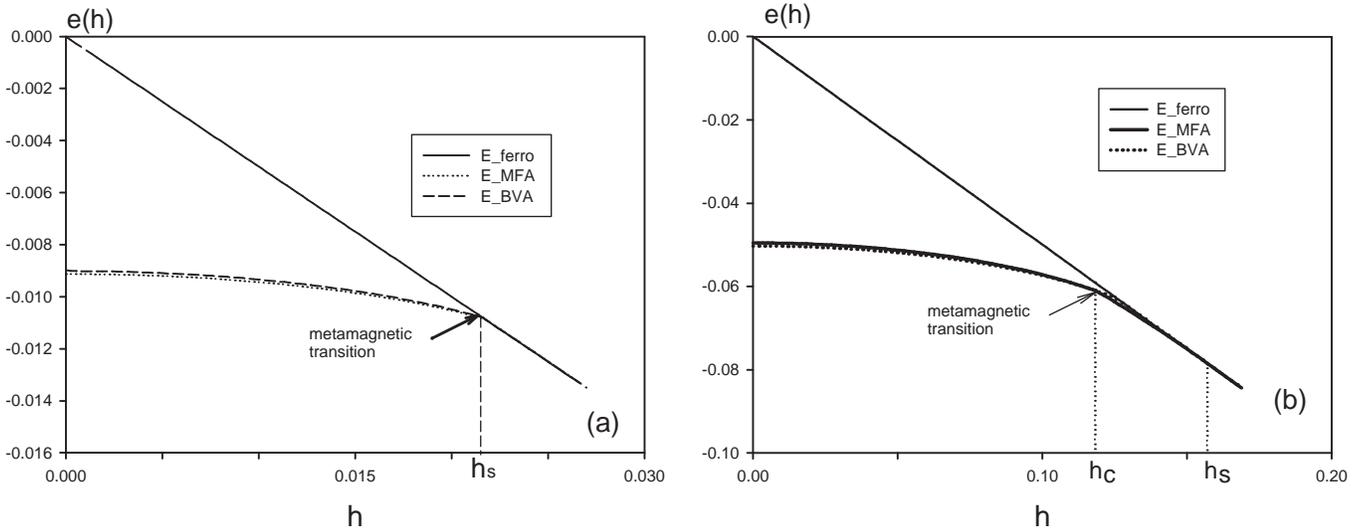}
\caption{The dependence of the ground state energy of the model (\ref{H1})
on magnetic field $e(h)$ for (a) $\protect\alpha=0.3$ and (b) $\protect\alpha%
=0.4$.}
\label{fig_2}
\end{figure}

\begin{figure}[tbp]
\includegraphics{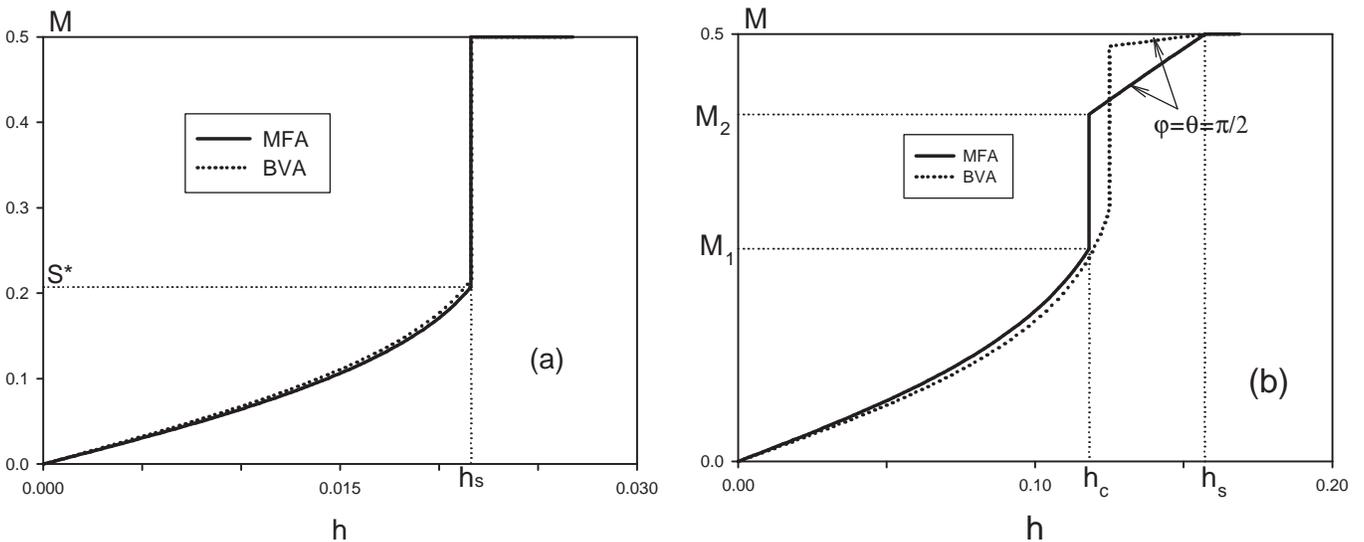}
\caption{The dependence of the magnetization of the model (\ref{H1}) on
magnetic field $M(h)$ for (a) $\protect\alpha=0.3$ and (b) $\protect\alpha%
=0.4$.}
\label{fig_3}
\end{figure}

\begin{figure}[t]
\includegraphics{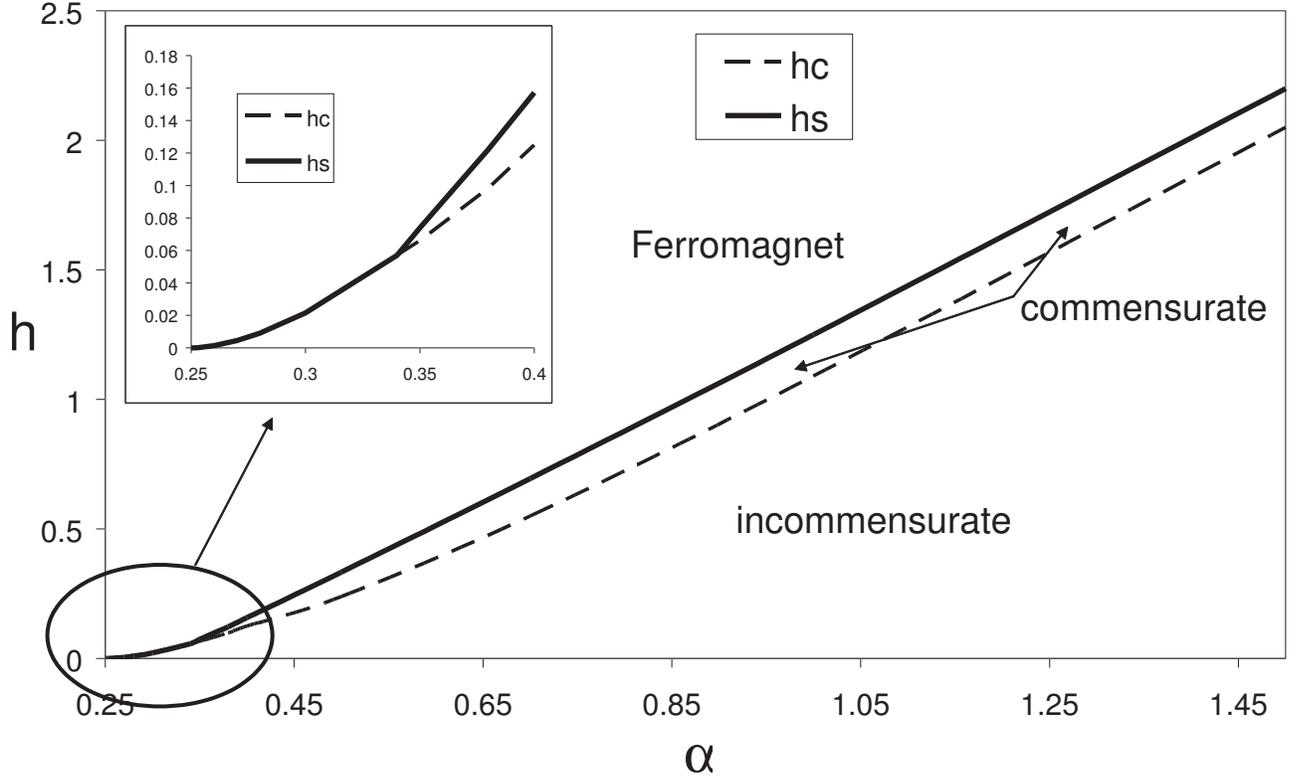}
\caption{The phase diagram of the model (\ref{H1}) in the MFA in the $(h,%
\protect\alpha)$ plane.}
\label{fig_4}
\end{figure}

We calculate the magnetization for the model (\ref{H1}) using the MFA at $%
h>0 $. In this case there are two variational parameters, $\varphi $ and $%
\theta $. We show the ground sate energy per site $e$ as a function of $h$
for two values of $\alpha $ in Fig.\ref{fig_2} together with the energy of
the ferromagnetic state $e_{F}=-\frac{h}{2}$. The saturation field $h_{s}$
is determined by the crossing between the variational energy $e(h)$ and $%
e_{F}$. The dependence $e(h)$ shown in Fig.\ref{fig_2} demonstrates two
different types of the behavior. For $\alpha \lesssim 0.38$ (Fig.\ref{fig_2}%
a) $e(h)$ and $e_{F}(h)$ have different slopes at the crossing point. At $%
h=h_{s}$ the magnetization jumps from $M^{\ast }=\left. \frac{\partial e}{%
\partial h}\right| _{h=h_{s}}$ to $\frac{1}{2}$ as it is shown in Fig.\ref
{fig_3}a. It means that maximum in Eq.(\ref{hc}) is reached for macroscopic $%
n$ and the metamagnetic transition takes place. The dependence of the
saturated field on $\alpha $ is shown in Fig.\ref{fig_4}. This dependence is
consistent with those obtained in \cite{Aligia} by using numerical
diagonalizations of finite chains. The magnetization curve at $0<\gamma \ll 1
$ can be found analytically from the self-consistency equations (\ref{ntq}).
The energy per site $e(M)$ at $M\ll 1/2$ has a form
\begin{equation}
e(M)=-1.585\gamma ^{\frac{12}{7}}+3.69\gamma ^{\frac{10}{7}}M^{2}+O(\gamma ^{%
\frac{12}{7}})
\end{equation}

Performing the Maxwell construction we obtain
\begin{eqnarray}
M^{\ast } &=&0.429\gamma ^{\frac{2}{7}}  \nonumber \\
h_{s} &=&-3.17\gamma ^{\frac{12}{7}}  \label{hs}
\end{eqnarray}

As follows from Eq.(\ref{hc}) $h_{s}$ in Eq.(\ref{hs}) is the binding energy
per magnon of multimagnon complex. It is interesting to compare this value
with the binding energy of two-magnon complex which is $-72\gamma ^{3}$ at $%
0<\gamma \ll 1$ \cite{Chubukov,KO}. It is evident that interaction of a
macroscopic number of magnons essentially lowers the binding energy.

The magnetization $M(h)$ at $0<\gamma \ll 1$ is
\begin{eqnarray}
M(h) &=&0.136\gamma ^{-\frac{10}{7}}h,\text{ \ }h<h_{s}  \nonumber \\
M(h) &=&\frac{1}{2},\text{ \ }h>h_{s}  \label{M(h)}
\end{eqnarray}

According to Eq.(\ref{M(h)}) the susceptibility $\chi $ is
\begin{equation}
\chi =0.136\gamma ^{-\frac{10}{7}}
\end{equation}

As follows from Eq.(\ref{M(h)}) and as shown on Fig.\ref{fig_3} the
magnetization is a linear function of $h$ at $h\rightarrow 0$. This
testifies the absence of the gap in the spectrum. As it is noted before, the
subtle question about the gap in this model is still open. Numerical
calculations show $1/N$ behavior for a gap \cite{Cabra}, while the one-loop
renormalization group indicates so tiny exponentially small gap \cite{Itoi}
that it can not be observed numerically. The proposed version of the MFA
does not predict the dimerization and the gap for the model (\ref{H1}).
However, we are not sure that the mean-field approach is an adequate tool to
answer so subtle question.

For $\alpha \gtrsim 0.38$ the curve $e(h)$ (Fig.\ref{fig_2}b) has a cusp at
some critical magnetic field $h_{c}$ as a result of the jump of variational
parameters minimizing energy from $\varphi ,\theta <\frac{\pi }{2}$ to $%
\varphi =\theta =\frac{\pi }{2}$. At $h=h_{c}$ the magnetization jumps from $%
M_{1}=-\left. \frac{\partial e(h)}{\partial h}\right| _{h=h_{c}-}$to $%
M_{2}=-\left. \frac{\partial e(h)}{\partial h}\right| _{h=h_{c}+}$.
According to Eq.(\ref{Jxyz}) $e(h)$ at $h>h_{c}$ ($\varphi =\theta =\frac{%
\pi }{2}$) is the ground state energy of the Hamiltonian
\begin{equation}
H=-\sum_{n=1}^{N}(S_{n}^{z}S_{n+1}^{z}-\frac{1}{4})+\alpha \sum_{n=1}^{N}(%
\mathbf{S}_{n}\mathbf{S}_{n+2}-\frac{1}{4})-h\sum_{n=1}^{N}S_{n}^{z}
\label{Hising}
\end{equation}

The curve $e(h)$ in Fig.\ref{fig_2}b is tangent to $e_{F}(h)$ at the
crossing point. The saturation field $h_{s}$ and the magnetization $M(h)$
near $h_{s}$ are determined from the self-consistency equations (\ref{ntq})
with $\varphi =\theta =\frac{\pi }{2}$
\begin{eqnarray}
h_{s} &=&-\frac{E_{2}}{2}  \nonumber \\
M &=&\frac{1}{2}-\frac{2(1+\alpha )^{4}}{(1+2\alpha )(4\alpha ^{4}+15\alpha
^{3}+11\alpha ^{2}-1)}(h_{s}-h)  \label{hsM}
\end{eqnarray}
where
\begin{equation}
E_{2}=-\frac{4\alpha ^{2}+2\alpha -1}{1+\alpha }  \label{E22}
\end{equation}

It is interesting to note that $E_{2}$ given by this equation is the exact
energy of the bound two-magnon state with a momentum $q=\pi $ of the
Hamiltonian (\ref{Hising}). The wave function of this state has a form
\begin{equation}
\left| \psi \right\rangle =\sqrt{\frac{(1-e^{-2/\xi })}{N}}%
\sum_{n,l}(-1)^{n+l}e^{-l/\xi }S_{n}^{-}S_{n+2l+1}^{-}\mid \uparrow \uparrow
\uparrow ...\uparrow \rangle  \label{psi}
\end{equation}
with a correlation length
\begin{equation}
\xi ^{-1}=\ln \frac{1+\alpha }{\alpha }  \label{xi}
\end{equation}

Remarkable, that this function is the exact one of the two-magnon state of
the Hamiltonian (\ref{H1}) as well because
\begin{equation}
\sum (S_{n}^{x}S_{n+1}^{x}+S_{n}^{y}S_{n+1}^{y})\left| \psi \right\rangle =0
\end{equation}

The state (\ref{psi}) has the lowest energy in two-magnon sector of the
Hamiltonian (\ref{H1}) at $\alpha >0.38$ \cite{Chubukov}.

Thus, the maximum in Eq.(\ref{hc}) is reached at $n=2$ if $e(h)$ has the
form similar to that shown in Fig.\ref{fig_2}b. This fact confirms the
observation based on finite-size results \cite{Cabra} that in the region of $%
\alpha $ rather far from $\alpha =\frac{1}{4}$ only the flipping of pairs of
spins participate in the magnetization process near the saturation.

The dependence $M(h)$\ for two typical cases is shown in Fig.\ref{fig_3}. At
$\alpha >0.38$ the MFA gives the saturation field exactly, though the linear
behavior of $M(h)$\ in Eq.(\ref{hsM}) is in a contrast with the expected
square-root critical dependence.

It was determined above that the presence of the cusp in the dependence $%
e(h) $\ is the result of the jump of variational parameters. At $h<h_{c}$\
the angle $\varphi $\ is incommensurate while at $h>h_{c}$\ it corresponds
to the commensurate phase. Therefore, the magnetic field $h_{c}$\ can be
identified with the critical field at which the incommensurate-commensurate
transition takes place. The phase diagram of the model (\ref{H1}) in the MFA
is shown in Fig.\ref{fig_4}. The commensurate phase lies between the
incommensurate and\ the ferromagnetic phases.

In the limit $\alpha \rightarrow \infty $ the Hamiltonian (\ref{H1})
describes two independent Heisenberg chains. The numerical solution of the
MFA equations indicates that in this limit the width of the commensurate
phase shrinks as $1/\alpha $. In the limit $\alpha \rightarrow \infty $ the
pitch angle $\varphi $ tends to $\pi /2$\ for all values of $h$, while the
canted angle $\theta $\ changes from $\theta =0$\ at $h=0$\ to $\theta =\pi
/2$\ at $h=h_{s}$. This means that the MFA correctly describes the limit $%
\alpha \rightarrow \infty $ showing the incommensurate phase at $0<h<h_{s}$.

The MFA shows the jump in magnetization at some critical field $h_{c}$. We
are not sure whether a true magnetization curve has (or not) the jump.
However, we believe that the jump obtained in our approximation testifies
the singular behavior of the magnetization at some critical field. A
plausible reason of such singularity consists in the following physical
picture. As was shown above at $\alpha >0.38$ two magnons attract one
another and form the bound two-magnon state (\ref{psi}) with the correlation
length $\xi $ (\ref{xi}). When the magnetic field is slightly lower than the
saturated field $h_{s}$ and a number of magnons is small, the ground state
can be represented with high accuracy as a product of\ the bound magnon
pairs, weakly interacted with each other. The bound pairs of magnons start
to feel each other when the mean distance between bound pairs becomes of the
order of the correlation length of the pair $\xi $. This happens at some
critical magnon density $n_{c}\sim \xi ^{-1}$. For larger density $n>n_{c}$
one can not consider magnon pairs as independent quasi-particles and the
above picture is destroyed. Certainly, these speculations can be well
justified in the case $\alpha \gg 1$ only, when the critical density is
small $n_{c}\sim 1/\alpha $.

Possible\emph{\ }singularity in the magnetization is related to the
magnetization curve in $Rb_{2}Cu_{2}Mo_{3}O_{12}$. The experimental
magnetization at $T=2.6K$ demonstrates a sufficiently sharp increase up to $%
M\simeq 0.4$ at $B\simeq 14$ Tesla followed by a gradual increase to the
saturation. In the light of above mentioned one can assume that the sharp
change in the behavior of $M(h)$ at $B\simeq 14T$ is connected with the
transition from the incommensurate regime to the commensurate one. Taking
the values of $J_{1}$\ and $\alpha $ estimated for $Rb_{2}Cu_{2}Mo_{3}O_{12}$%
\ ($\alpha \simeq 0.4,J_{1}\simeq -140K$) \cite{Hase} we obtain $B_{c}=13.6T$%
.

\section{Boson variational approach}

Another technique which can be used for an approximate analysis of the model
(\ref{H1}) is the boson variational approach (BVA) proposed in Ref.\cite
{boson}. This approach is based on Agranovich-Toshich boson representation
of spin $s=1/2$ operators \cite{Agranovich}:
\begin{eqnarray}
S_{n}^{z} &=&-\frac{e^{i\pi b_{n}^{+}b_{n}}}{2}  \nonumber \\
S_{n}^{-} &=&\frac{1-e^{i\pi b_{n}^{+}b_{n}}}{2\sqrt{b_{n}^{+}b_{n}}}%
b_{n}^{+}  \nonumber \\
S_{n}^{+} &=&b_{n}\frac{1-e^{i\pi b_{n}^{+}b_{n}}}{2\sqrt{b_{n}^{+}b_{n}}}
\label{spin-boson}
\end{eqnarray}
where $b_{n}^{+}$ are the Bose operators. This transformation preserves all
commutation relations for the spin operators and does not involve
`nonphysical states', because all states with odd number of bosons
correspond to the same spin state $\left| \downarrow \right\rangle $ and the
states with even number of bosons correspond to the spin state $\left|
\uparrow \right\rangle $. The latter implies that the transformation (\ref
{spin-boson}) is not unitary and each energy level of spin model corresponds
to infinite degenerated level of boson Hamiltonian. However, this
transformation allows to estimate the ground state energy of the spin model.

The Hamiltonian of the considered spin model rewritten in the bosonic form
is treated by the variational function in the form \cite{boson}:
\begin{equation}
\left| \Psi \right\rangle =\exp \left( \frac{1}{2}\sum_{i,n}\Lambda
(n)b_{i}^{+}b_{i+n}^{+}\right) \left| 0_{b}\right\rangle  \label{psi-boson}
\end{equation}
where $\left| 0_{b}\right\rangle $ is the boson vacuum state corresponding
to the ferromagnetic spin state with all spins pointing up and the function $%
\Lambda (n)$ is chosen by the condition of minimum of the total energy.
Therefore, in contrast to spin-wave theory, this approach is variational.
The procedure of calculation of the variational energy with the wave
function (\ref{psi-boson}) and energy minimization over function $\Lambda
(n) $ was developed in \cite{boson}, where the approach was successfully
applied to construction of the ground state phase diagram of the frustrated
2D Heisenberg model.

We have applied the above approach to the rotated spin Hamiltonian (\ref
{Hrot}) and the rotation angles $\varphi $ and $\theta $ was used as
variational parameters, as was in the MFA. The contribution to the energy of
the parts $H_{2}$ and $H_{3}$ of the Hamiltonian (\ref{Hrot}) in this
approach is also zero as in the MFA. We do not present here cumbersome
details of calculations, because they are identical to those in Ref.\cite
{boson} and give only the final results.

The approach shows very similar behavior of the ground state energy as in
the Jordan-Wigner MFA. That is, for small $0<\gamma \ll 1$ both approaches
give the same critical exponent for the ground state energy and for the
parameter $\varphi $. Moreover, the numerical estimates of the ground state
energy in both approaches are also very close (see Fig.\ref{fig_1}). The
behavior of the magnetization curve in the BVA are also very similar to that
in the MFA (see Fig.\ref{fig_3}). The perfect coincidence of physical
properties of the model predicted by the MFA and the BVA looks somewhat
surprising, because of different nature of these approaches.

We note, that the potential advantage of the boson variational approach
consists in its applicability to any dimension lattices and possible
modification to any spin value $S$.

\section{Scaling estimate of the critical exponent near the transition point
$\protect\alpha =1/4$}

As was shown in previous sections the MFA and the BVA gives the ground state
energy $E(\gamma )\sim \gamma ^{12/7}$ near the transition point $\alpha =1/4
$. Since both approaches used are variational, the value $\beta =\frac{12}{7}
$\ is upper bound for the critical exponent $\beta $. This implies an
important and strict fact that quantum fluctuations definitely change the
classical critical exponent $\beta =2$.

Nevertheless, there are some reasons to believe that true value of the
critical exponent is $\beta =\frac{5}{3}$. The arguments in favor of this
conjecture are based on the following consideration with use of the results
of numerical calculations of finite-size chains \cite
{Tonegawa89,Richter,future}.

The ground state of the model (\ref{H1}) is the singlet at $\alpha >\frac{1}{%
4}$ for any\ even $N$. For cyclic chain there are number of level crossings
of two lowest singlets with momenta $q=0$ and $q=\pi $. For example, for $%
N=4k$ there are $k-1$ crossing points $\frac{1}{4}<\alpha _{1}(N)<\alpha
_{2}(N)<\ldots <\alpha _{k-1}(N)$ \cite{Tonegawa89,Richter}. This fact means
that in the thermodynamic limit the ground state is at least two-fold
degenerated at $\alpha >\frac{1}{4}$. Let us determine the dependence of $%
\alpha _{1}$ (actually, $\gamma _{1}=\alpha _{1}-\frac{1}{4}$) on $N$. It is
known this dependence defines the scaling parameter in one-dimensional
models, where the crossings of two ground state levels occur. At $\alpha =%
\frac{1}{4}$ $(\gamma =0)$ the ground state in the sector $S=0$ has momentum
$q=\pi $, while the first excited state has $q=0$ and the excitation energy $%
\Delta E\sim N^{-4}$ \cite{future}. The first order correction to the energy
of these states in $\gamma $ is
\begin{equation}
\delta E_{1(2)}(\gamma )=\gamma \sum_{n}\langle \Psi _{1(2)}\left| \mathbf{S}%
_{n}\cdot \mathbf{S}_{n+2}\right| \Psi _{1(2)}\rangle  \label{E1}
\end{equation}
where $\Psi _{1}$ and $\Psi _{2}$ are the wave functions of the ground state
and the first excited singlet at $\alpha =\frac{1}{4}$. These two states
have a spiral ordering at $N\rightarrow \infty $ with a period of spirals $N$
and $\frac{N}{2}$ respectively \cite{Tonegawa89,future}. According to this
fact the two-spin correlation functions at $N\rightarrow \infty $ are
\begin{eqnarray}
\langle \Psi _{1}\left| \mathbf{S}_{n}\cdot \mathbf{S}_{n+l}\right| \Psi
_{1}\rangle &=&\frac{1}{4}\cos \frac{2\pi l}{N}  \nonumber \\
\langle \Psi _{2}\left| \mathbf{S}_{n}\cdot \mathbf{S}_{n+l}\right| \Psi
_{2}\rangle &=&\frac{1}{4}\cos \frac{4\pi l}{N}
\end{eqnarray}

The accuracy of the above equations for $l=2$ is of the order of $O(N^{-3})$
\cite{Hamada,future}.

The value $\gamma _{1}$ is determined from the condition
\begin{equation}
\delta E_{1}(\gamma _{1})-\delta E_{2}(\gamma _{1})=\Delta E
\end{equation}
which gives
\begin{equation}
\gamma _{1}\sim N^{-3}
\end{equation}

Therefore, the scaling parameter of the model (\ref{H1}) in the vicinity of
the transition point $\alpha =\frac{1}{4}$ is $x=\gamma N^{3}$. It means
that the perturbation theory in $\gamma $ contains infrared divergencies and
the ground state energy has a form
\begin{equation}
E(\gamma )=\frac{\gamma }{N}f(x)  \label{E2}
\end{equation}
where the scaling function $f(x)$ at $x\rightarrow 0$ is given by the first
order correction (\ref{E1}). In the thermodynamic limit ($x\rightarrow
\infty $) the behavior of $f(x)$ is generally unknown, but the condition $%
E\sim N$ at $N\rightarrow \infty $ requires
\begin{equation}
E(\gamma )\sim N\gamma ^{\beta }  \label{E3}
\end{equation}

According to Eqs.(\ref{E2})-(\ref{E3}) $\beta =\frac{5}{3}$. This value is
close to $\beta =\frac{12}{7}$ which indicates high quality of the
variational approaches used. A possible reason of a discrepancy between the
variational and scaling estimates of the exponent may be related to the fact
that in the variational approaches the terms $H_{2}$ and $H_{3}$ in Eq.(\ref
{Hrot}) are irrelevant. \emph{\ }It can be also expected that the true
dependences in Eqs.(\ref{Egamma}),(\ref{hs}) at small $0<\gamma \ll 1$ are
\begin{equation}
\varphi \sim \gamma ^{\frac{1}{3}},\qquad M^{\ast }\sim \gamma ^{\frac{1}{3}%
},\qquad h_{s}\sim \gamma ^{\frac{5}{3}},\qquad \chi \sim \gamma ^{-\frac{4}{%
3}}
\end{equation}

\section{Summary}

We have studied the frustrated spin-$\frac{1}{2}$ Heisenberg chain with the
NN ferromagnetic and the NNN antiferromagnetic exchange interactions using
two different variational approximations: the MFA and the BVA. The first
step of both approaches consists in rotation of the coordinate system on the
pitch angle $\varphi $ and the canted angle $\theta $, which are not equal
to their classical values and used as variational parameters of the
approaches. Then, in the MFA the rotated spin Hamiltonian is mapped into the
model of interacting spinless fermions by means of the Jordan-Wigner
transformation. The latter is treated in the mean-field approximation with
the inclusion of superconductor-like correlations. Within the BVA we use the
Agranovich-Toshich boson transformation of spin operators and the
variational treatment of the bosonic Hamiltonian. A remarkable fact, despite
the difference of these approaches they give quantitatively close results.

The variational approaches used allowed us to estimate the critical exponent
of the ground state energy $\beta $ in the vicinity of the transition point
from the ferromagnetic state to the singlet one. Both approaches give $\beta
=\frac{12}{7}$ which differs from the classical value $\beta =2$. Since
approaches used are variational, we have established an important and strict
fact that quantum fluctuations definitely change the classical critical
exponent. Using the results of finite-size calculations we presented also
some scaling arguments in favor of the critical exponent $\beta =\frac{5}{3}$%
. This value is close to $\beta =\frac{12}{7}$\ which indicates high quality
of the variational approaches used.\emph{\ }

The behavior of the magnetization curve is different in parameter regions $%
\alpha \lesssim 0.38$\ and $\alpha \gtrsim 0.38$. In the region $\alpha
\lesssim 0.38$\ the metamagnetic transition to saturation takes place. At $%
\alpha \gtrsim 0.38$ the magnetization increases with $h$\ until the field
reaches the critical value $h_{c}$,\ where the magnetization jumps from $%
M_{1}$\ to $M_{2}<\frac{1}{2}$. This magnetization jump accompanies the jump
of the pitch angle $\varphi $ and canted angle $\theta $. At $h<h_{c}$\ both
angles are incommensurate, while at $h>h_{c}$\ they correspond to the
commensurate phase $\varphi =\theta =\pi /2$. Therefore, we associate this
jump with the incommensurate-commensurate transition induced by the magnetic
field.

We believe that our approaches correctly predict the existence of the
incommensurate-commensurate transition at some critical field $h_{c}$ and $%
\alpha >0.38$. This transition must accompany some singular behavior of the
magnetization curve, though we are not sure whether the true magnetization
has the jump at $h=h_{c}$.

Both approximations used yield the exact value of the saturation field at $%
\alpha >0.38$, though the field dependence near the saturation does not
correspond to the expected universal square-root behavior. We note that the
magnetization behavior near the saturation in the region of sufficiently
large value of $\alpha $\ is described by the model (\ref{Hising}), which
looks simpler than the model (\ref{H1}). In this respect an accurate study
of the model (\ref{Hising}) is of a particular interest.

We note that the magnetization curve in Fig.(\ref{fig_3}) resembles (apart
from the magnetization jump) that observed in $Rb_{2}Cu_{2}Mo_{3}O_{12}$.
This resemblance allows us to provide a qualitative explanation of the
peculiarity of the magnetization process in this compound. The experimental
curve is characterized by a sharp change of the susceptibility at the
magnetic field $B\simeq 14$ Tesla. We assume that this peculiarity in the
magnetization behavior is related to the crossover between the
incommensurate and the commensurate states.

This work was supported under RFBR Grant No 03-03-32141.

\end{document}